\documentclass[dvips]{article}
\usepackage{wasysym}
\usepackage{icrctc07}

\title{Observations of Very High Energy Gamma-Rays during Moonlight and Twilight with the MAGIC Telescope}
\shorttitle{VHE $\gamma$-rays during Moonlight and Twilight with the MAGIC Telescope}

\authors{Javier Rico$^{1}$, Emma de O\~na-Wilhelmi$^{1}$,Juan
Cortina$^{1}$, Eckart Lorenz$^{2}$ for the MAGIC Collaboration}
\shortauthors{J. Rico et al.}
\afiliations{$^1$Institut de Fisica d'Altes Energies, Barcelona, Spain\\ 
$^2$Max-Planck-Institut f\"ur Physik, M\"unchen, Germany}
\email{jrico@ifae.es, emma@ifae.es}

\abstract{We study the capability of the MAGIC telescope to observe under
moderate moonlight. TeV $\gamma$-ray signals from the Crab nebula were
detected with the MAGIC telescope during periods when the Moon was
above the horizon and during twilight. This was accomplished by
increasing the trigger discriminator thresholds. No change is
necessary in the high voltage settings since the camera PMTs were
especially designed to avoid high currents. We characterize the
telescope performance by studying the effect of the moonlight on the
$\gamma$-ray detection efficiency and sensitivity, as well as on the
energy threshold.}

\begin{document}
\maketitle
\section{Introduction}

Ground-based searches for very high energy (VHE) $\gamma$-ray emission
from celestial objects are normally carried out by so-called imaging
air Cherenkov telescopes (IACT) during clear, moonless nights. If such
a strict requirement is relaxed to allow observations under moderate
moonlight or twilight, an increase of the duty cycle to 18$\%$ (from
$\sim$1000 to $\sim$1500 hours of observations per year) is
possible. VHE observations under moonlight have been tested in the
past~\cite{Pare,Chantell,Kranich} but with solutions that were too
expensive, time consuming and not efficient in terms of energy
threshold and sensitivity. The MAGIC IACT~\cite{Lorenz} has been
designed to carry out observations also during moderate
moonlight. This places MAGIC in a prominent position, in particular
for the study of variable sources as well as in multi-wavelength
campaigns together with other instruments. In this paper we describe
the technical innovations and analysis changes that allow observations
in the presence of the Moon with MAGIC.

Traditionally, PMTs are operated with amplification gains around
$10^6-10^7$ which, under moonlight, generate continuous (direct)
currents (DCs) that can damage the last dynode, resulting in rapid
ageing of the PMT. MAGIC PTMs run at a gain of about
3$\times10^4$~\cite{Performance}. In order to also detect single
photoelectrons (phe) the PMT signal is fed to an AC-coupled fast, low
noise preamplifier to raise the combined gain to about $10^6$. The PMT
analog signal is transmitted over an optical fiber, converted into an
electrical pulse and split into two branches, one of which enters a
discriminator (DT), which issues a digital signal whenever
the pulse exceeds a given threshold.

\begin{figure}[t]
\begin{center}
\includegraphics[width=0.45\textwidth]{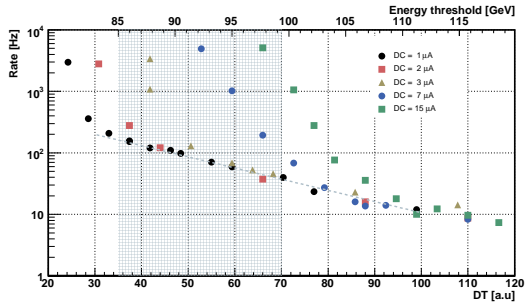}
\caption{Trigger rate as a function of the discriminator
threshold for four neighboring pixels configuration and different
camera illuminations. The shaded area shows the range used for MAGIC
regular observations (dark and under moonlight). The dashed line shows
the linear regime. The upper axis shows the corresponding energy
threshold (after image cleaning) for observations at zenith angles
between 20$^\circ$ and 30$^\circ$ as deduced in
Section~\ref{sec:energy}.}
\label{rate}
\end{center}
\end{figure}

The increase of the background light due to the presence of the Moon
depends on various factors including Moon and source zenith angles,
Moon phase, angular distance to the Moon and atmospheric
conditions. At around 25$^\circ$ away from the Moon the direct
scattered moonlight approaches a constant level below that of the
night sky light background. Figure~\ref{rate} shows the dependence of
the trigger rate (for a four neighboring pixels configuration) on the
DT settings, for different illumination conditions (which produce
different anode currents in the camera). The light of night sky (LONS)
is responsible for the steep increase at low DT values (at $\sim$30
a.u. in the case of dark observations). At higher DT values the rate
is caused mostly by Cherenkov showers. The telescope operates at the
minimum possible DT for which the contribution of accidental triggers
is negligible. For extragalactic regions (DC=$1\mu$A) the DTs are
generally set to 30 a.u., which corresponds to a pulse charge of
8-10~phe. Higher DT values are needed to keep the trigger rate below
the limit of the DAQ system (500 Hz) for observations during twilight
and moonlight. We restrict MAGIC observations to a maximum DC of
8~$\mu$A. This permits observations in the presence of the Moon until
(since) 3-4 days before (after) full Moon, for an angular distance to
the Moon greater than 50$^\circ$~\cite{Performance}.

\section{Observations and Data Analysis} %

To characterize the response of the telescope under moonlight, we
observed the Crab nebula at different light conditions (including dark
observations used as reference) between January and March 2006, in the
ON/OFF mode. Two data sets, one with zenith angle between 20$^\circ$
and 30$^\circ$, and a second one between 30$^\circ$ and 40$^\circ$,
were acquired and analyzed separately. Depending on the different
moonlight levels, the resulting anode currents ranged between 1 and
6~$\mu$A. Correspondingly, the DT was varied between 35 and 65 units.
The acquired data were processed by the standard MAGIC analysis
chain~\cite{signal}. The images were cleaned using absolute tail and
boundary cuts at 10 and 5~phe, respectively. Quality cuts based on the
trigger and after-cleaning rates were applied in order to remove bad
runs. The shower images were parameterized using the Hillas
parameters~\cite{Hillas} SIZE, WIDTH, LENGTH, DIST, CONC and ALPHA,
combined (except for the latter) into an adimensional variable
(HADRONNESS) for $\gamma$/hadron separation by means of a Random
Forest classification algorithm~\cite{Breiman}. The signal region was
defined by the cuts HADRONNESS$<$0.15 and ALPHA$<$8$^\circ$.

\section{Results} %

\label{sec:energy}
\begin{figure}[t]
\begin{center}
\includegraphics[width=0.45\textwidth]{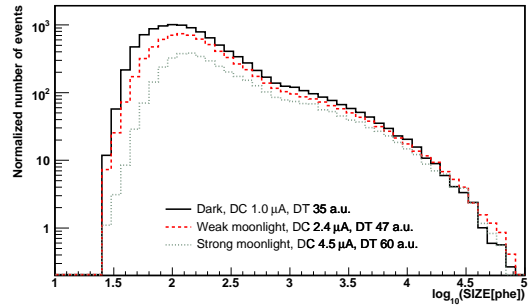}
\caption{Distributions of SIZE before analysis cuts for three
Crab nebula samples acquired under different light conditions. The
histograms have been normalized to a common observation time. Note
that the distributions are completely dominated by hadronic events
($\sim 99\%$).}
\label{dist_size}
\end{center}
\end{figure}

\begin{figure}[t]
\begin{center}
\includegraphics[width=0.22\textwidth]{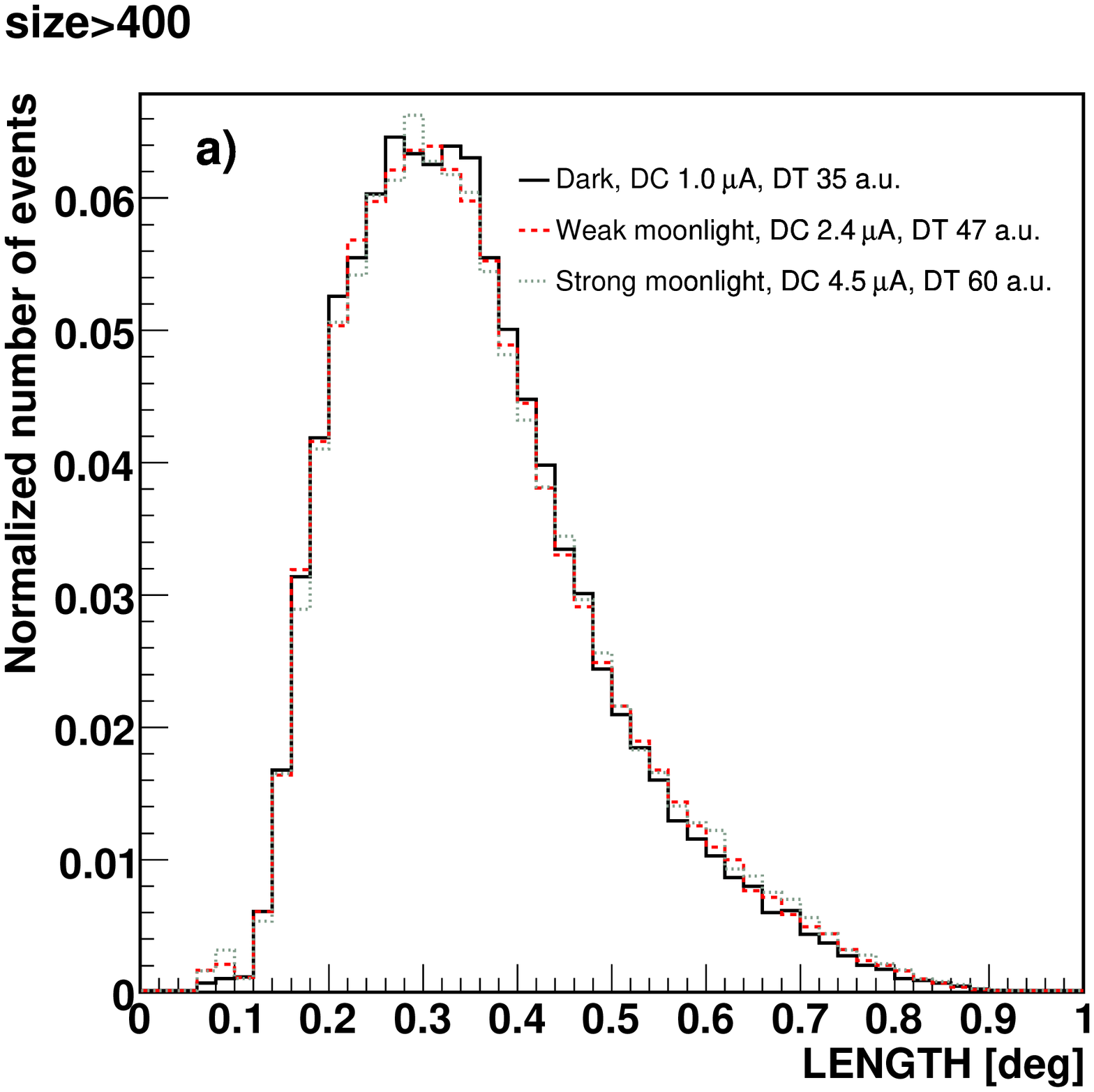}
\includegraphics[width=0.22\textwidth]{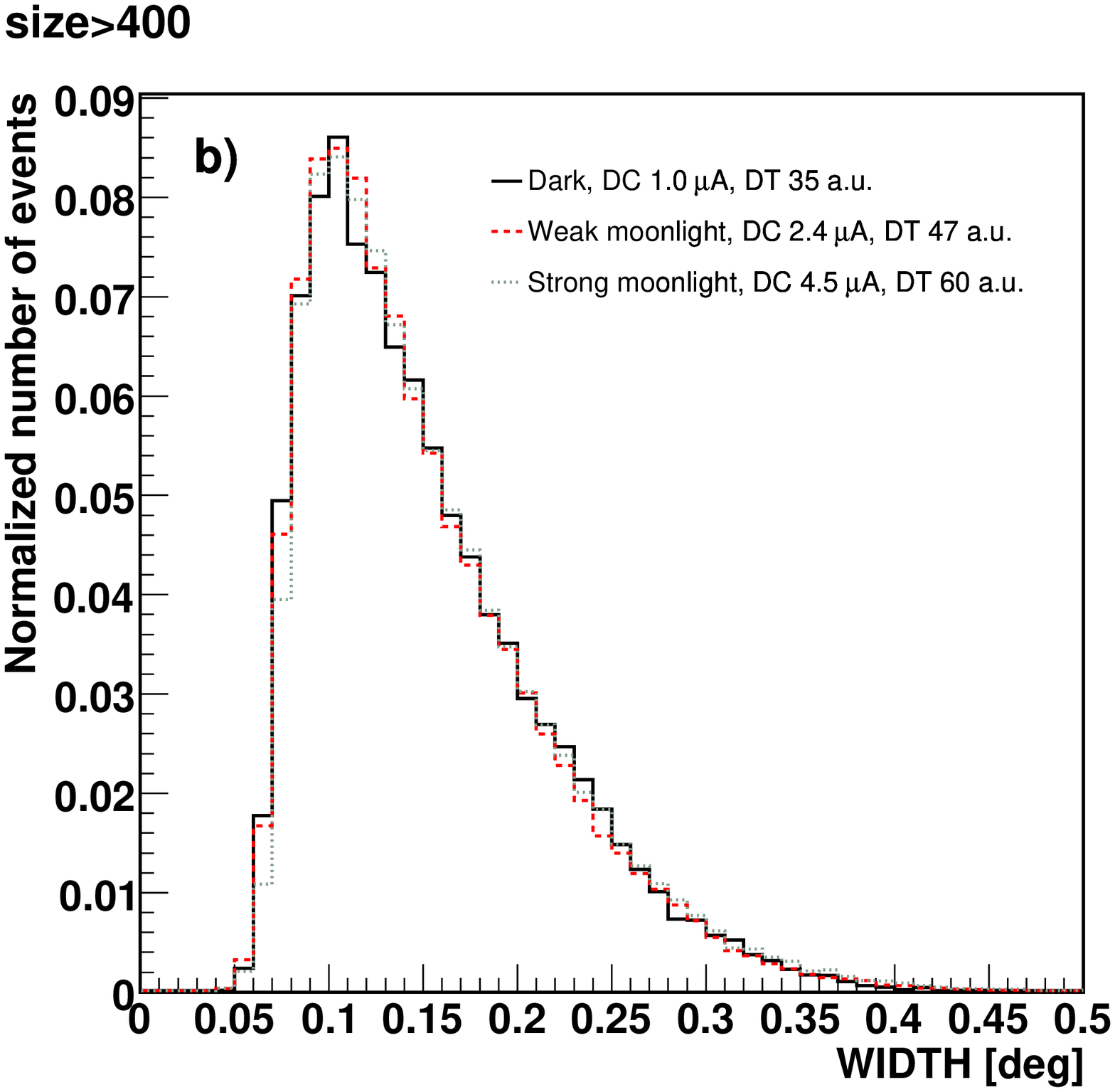}
\includegraphics[width=0.22\textwidth]{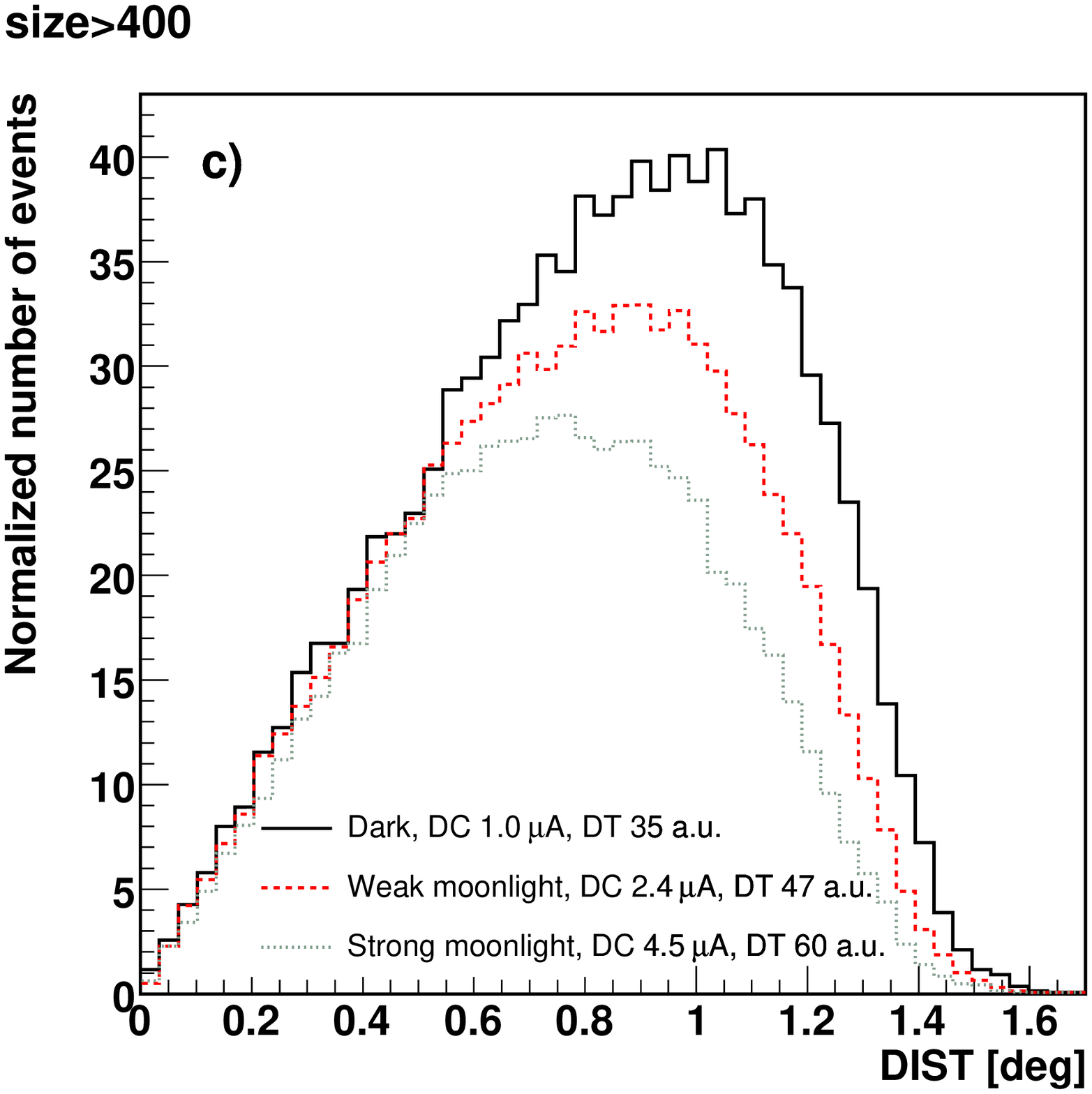}
\includegraphics[width=0.22\textwidth]{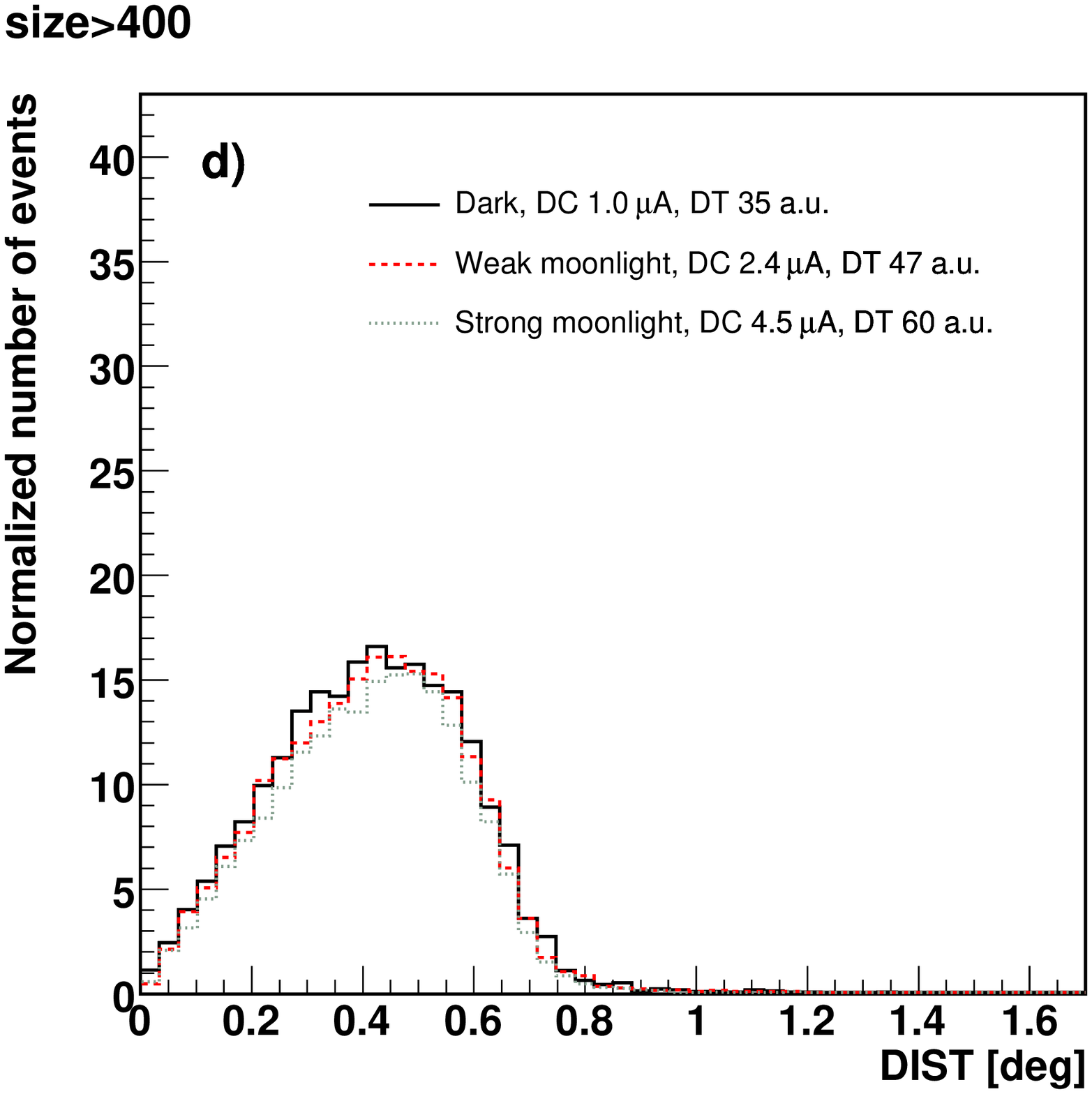}
\caption{Distributions of LENGTH (a), WIDTH (b) and DIST for
all recorded events (c) and for images fully contained in the inner
camera (d) for SIZE$>$400 phe. Three Crab nebula samples acquired
under different moonlight conditions. The histograms are normalized to
a unit area in (a) and (b) and to a common observation time in (c) and
(d).}
\label{dist_temp}
\end{center}
\end{figure}

As we increase the DT levels to counteract accidental triggers, one
depletes the SIZE distribution of shower candidates, as expected,
mostly at low values. However, we find that a substantial number of
showers with SIZE up to $10^4$~phe, i.e.\ those well above the trigger
level (which is around 50 phe), are also suppressed (see
Figure~\ref{dist_size}). On the other hand, the LENGTH, WIDTH and CONC
distributions above 200~phe do not show significant differences (See
Figure~\ref{dist_temp}) and therefore we do not expect $\gamma$/hadron
separation to degrade due to the presence of the Moon. Below 200 phe
the distributions are distorted by the different trigger threshold. 

The supressed high-SIZE showers are those at high DIST (see
Figure~\ref{dist_temp}c), corresponding to a high impact parameter,
where a significant fraction of the light falls outside the trigger
area (1$^\circ$ radius around the camera center). In some cases, the
fraction of the shower image contained in the trigger area does not
exceed the increased threshold for at least 4 neighboring pixels, as
required for a trigger. A confirmation of this hypothesis was obtained
by two different tests. First, the DIST distributions for events fully
contained in the trigger region were compared
(Figure~\ref{dist_temp}d). In such a case we find similar
distributions. The second test was performed by observing Crab in dark
conditions (DC$\sim$1.1~$\mu$A), but with increased DTs. In this case
we found similar inefficiencies as those shown in Figure
\ref{dist_temp}c). Therefore we can conclude that the change of the DIST
distribution is not related to the mean DC current (i.e. with the
camera illumination) but only to the DT level.

These results show that moonlight does not distort the images from
Cherenkov showers and therefore the analysis based on the Hillas
parameters does not have to be adapted for data acquired under
moonlight, and in particular the $\gamma$/hadron separation power is
not reduced for this kind of observations. In addition, the
differences that we find in the event rates and the DIST distributions
are exclusively due to the fact that the DTs were increased to keep a
low rate of accidental events, together with the fact that the trigger
area does not span the whole camera. With a faster DAQ system one
could have a fix DT level and deal with the extra noisy events
introduced by the moonlight during the offline analysis.

\begin{figure}[t]
\begin{center}
\includegraphics[width=0.45\textwidth]{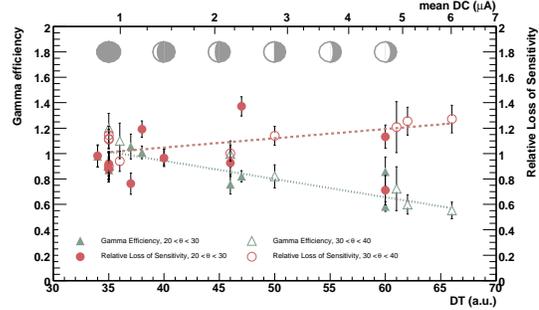}
\caption{Relative $\gamma$-ray detection efficiency (green triangles, left
axis) and sensitivity (red circles, right axis) as a function of DT
(SIZE$>$ 400~phe), for zenith angle bins $[20^\circ,30^\circ]$ (filled
markers) and $[30^\circ,40^\circ]$ (empty markers) measured from Crab
nebula observations.}
\label{resultsdata}
\end{center}
\end{figure}

\begin{figure}[t]
\begin{center}
\includegraphics[width=0.45\textwidth]{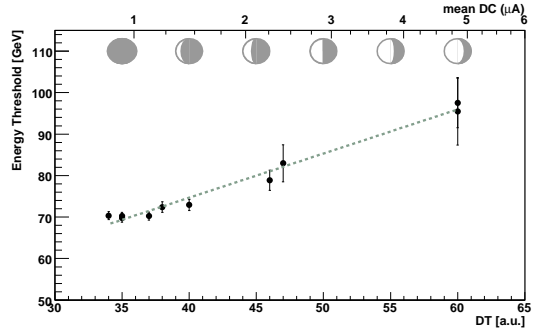}
\caption{Energy threshold after image cleaning
as a function of DT obtained from MC simulated $\gamma$-ray events
(for zenith angle between 20$^\circ$ and 30$^\circ$). The top axis
shows the typical mean DC for a chosen DT value.}
\label{threshold}
\end{center}
\end{figure}

For this study the dependence of the telescope response on the level
of Moon illumination will be parameterized as a function of DT. As is
shown in Figure~\ref{dist_size}, there is a reduction in the
collection area over a wide range of energies. We use Crab nebula
observations to parametrize the efficiency of detecting $\gamma$-rays,
for every DT and SIZE, relative to the values for dark
observations. These values are used during the off-line analysis
together with the MC simulation with standard DTs to calculate the
correct collection areas. For this, the observations of the Crab
nebula are divided into different samples according to the observation
date, SIZE and DT values. For each of the samples we get a measurement
of the $\gamma$-ray rate ($R$) and sensivity ($s$). We find that, for
a given SIZE range, the dependences of $R/R_0$ and $s/s_0$ with the DT
are well described by linear functions $(R/R_0,s/s_0 = \left(1 -
S_{R,s} \left(DT - DT_0\right)\right))$, where $R_0$, $s_0$ and $DT_0$
are the rate, sensitivity and DT values, respectively, for the dark
observation case. $S_{R,s}$ are let free during the fit and are
referred to as efficiency/sensitivity loss rate, respectively.  The
results for SIZE$>$400 phe and the two considered zenith angle samples
($[20^\circ,30^\circ]$ and $[30^\circ,40^\circ]$) are shown in
Figure~\ref{resultsdata}. The fit parameters obtained for both zenith
angles are compatible within statistical errors. This allows us to
perform a combined fit for both samples. For SIZE$>$400~phe we obtain
the following expression for the $\gamma$-ray efficiency and
sensitivity:
\begin{eqnarray*}
R/R_0  &=& 1 - (1.41\pm 0.32)\times10^{-2}(DT-DT_0) \\
s/s_0 & = & 1 + (6.3\pm 1.6)\times10^{-3}(DT-DT_0)
\label{eq:effloss}
\end{eqnarray*}

In order to understand the dependence of the $\gamma$-ray detection
efficiency on the energy we have performed the same study for four
bins of SIZE, namely [200,400], [400,800], [800,1600] and
[1600,6400]~phe, which roughly correspond to the energy ranges
[150,300], [300,600], [600-1000] and $>$~1000~GeV, respectively, for
low zenith angle. Up to SIZE=3000~phe we find a linear dependence that
can be parameterized by:
\begin{eqnarray*} 
S_R & = & (2.24\pm0.13)\times10^{-2} \\
    &   & - (7.2\pm 1.2)\times10^{-6} \textrm{ SIZE [phe]} \\
S_s & = & (1.63\pm0.14)\times10^{-2}  \\
    &   & - (7.4\pm 1.8)\times10^{-6}  \textrm{ SIZE [phe]}
\label{eq:efi_size}
\end{eqnarray*}
We tested the dependences of these results on different HADRONNESS and
ALPHA cuts, obtaining in all cases similar results.

Finally, it is important to understand the influence of the moonlight
on the energy threshold. We define the energy threshold as the peak of
the energy distribution of all events after image cleaning and before
analysis cuts. The dependence of the energy threshold is well
described by the following linear function (see
Figure~\ref{threshold}):
$$
\frac{E_\textrm{th}}{\textrm{GeV}} = (69.3\pm 0.4)+(1.06\pm 0.03)~(DT-DT_0) 
$$ 
This increase is relatively marginal, and it has to be noted again
that it is due to the increase of the DTs, and hence only indirectly
to the increase in the camera illumination.

We would like to thank the IAC for the excellent working conditions at
the Observatorio del Roque de los Muchachos in La Palma. 

\bibliography{icrc0557}

\begin{thebibliography}{1}

\bibitem{signal}
J.~{Albert} and {et al.}
\newblock {Signal Reconstruction for the MAGIC Telescope}.
\newblock {\em ArXiv Astrophysics e-prints}, December 2006.

\bibitem{Performance}
A.~{Armada}.
\newblock {Characterization and some applications of the anode current
  monitoring system of the MAGIC telescope}.
\newblock {\em Master Thesis}, 2005.

\bibitem{Breiman}
L.~{Breiman}.
\newblock {Random Forests}.
\newblock {\em Machine Learning}, 45:5--32, 2001.

\bibitem{Chantell}
M.~{Chantell} and {et al.}
\newblock {Gamma-Ray Observations in Moonlight with the Whipple Atmospheric
  Cherenkov Hybrid Camera}.
\newblock In {\em International Cosmic Ray Conference}, volume~2 of {\em
  International Cosmic Ray Conference}, pages 544--+, 1995.

\bibitem{Hillas}
A.~M. {Hillas}.
\newblock {Cerenkov light images of EAS produced by primary gamma}.
\newblock In F.~C. {Jones}, editor, {\em International Cosmic Ray Conference},
  volume~3 of {\em International Cosmic Ray Conference}, pages 445--448, August
  1985.

\bibitem{Kranich}
D.~{Kranich} and {et al.}
\newblock {TeV gamma-ray observations of the Crab and MKN 501 during moonshine
  and twilight}.
\newblock {\em Astroparticle Physics}, 12:65--74, October 1999.

\bibitem{Lorenz}
E.~Lorenz.
\newblock Status of the 17-m magic telescope.
\newblock {\em New Astron. Rev.}, 48:339--344, 2004.

\bibitem{Pare}
E.~{Pare} and {et al.}
\newblock {Image Shapes of Showers in UV and Visible Cherenkov Light}.
\newblock In {\em International Cosmic Ray Conference}, volume~1 of {\em
  International Cosmic Ray Conference}, pages 492--+, August 1991.

\end{thebibliography}
\bibliographystyle{plain}
\end{document}